\documentclass[a4paper]{jpconf}

\usepackage{graphicx}
\begin{document}
\title{D.C. Josephson transport by quartets and other Andreev resonances in superconducting bijunctions.}

\author{R. M\'elin$^1$, D. Feinberg$^1$, H. Courtois$^1$, C. Padurariu$^1$, A. Pfeffer$^2$, J.E. Duvauchelle$^2$, F. Lefloch$^2$, T. Jonckheere$^3$, J. Rech$^3$, T. Martin$^3$, B. Dou\c{c}ot$^4$}
\address{$^1$Universit\'e Grenoble-Alpes and CNRS, Institut N\'eel, Grenoble, France.}
\address{$^2$SPSMS/LaTEQS, UMR-E 9001, CEA-INAC and Universit\'e Grenoble-Alpes, Grenoble, France}
\address{$^3$Aix-Marseille Universit\'e, Universit\'e de Toulon, CNRS, CPT, UMR 7332, 13288 Marseille, France}
\address{$^4$Laboratoire de Physique Th\'eorique et des Hautes Energies,
  CNRS UMR 7589, Universit\'es Paris 6 et 7, 4 Place Jussieu, 75252 Paris
  Cedex 05}

\ead{denis.feinberg@neel.cnrs.fr}

\begin{abstract}
Bijunctions are three-terminal Josephson junctions where three superconductors are connected by a single weak link made of a metallic region or of quantum dots. Biasing two of the superconductors with commensurate voltages yields Andreev resonances that produce d.c. Josephson currents made of correlated Cooper pairs. For instance with applied voltages $(0, V, -V)$,  quartets formed by two entangled Cooper pairs are emitted by one reservoir towards the two others. Theory involving non-equilibrium Green's functions reveal the microsopic mechanism at play, e.g multiple coherent Andreev reflections that provide an energy-conserving and fully coherent channel. Recent experiments on diffusive Aluminum-Copper bijunctions show transport anomalies that are interpreted in terms of quartet resonances.

\end{abstract}

\section{Introduction}
Josephson junctions are weak links connecting two superconductors \cite{TINKHAM}. Cooper pairs crossing between the two condensates establish phase coherence \cite{JOSEPHSON}. The transport properties depend on the phase difference $\varphi$ between the two reservoirs. Zero voltage transport involves the d.c. Josephson current made of a steady flow of Cooper pairs, while at nonzero voltage, due to the relation $\frac{d\varphi}{dt}=\frac{2eV}{\hbar}$, pair transport manifests through the a.c. Josephson effect with oscillating Cooper pairs. In transparent enough $SNS$ junctions, transport occurs through  Andreev reflections at the $SN$ interfaces, forming Andreev electron-hole bound states at zero voltage. At nonzero voltage below the superconducting gap $\Delta$, quasiparticle transport can occur through multiple Andreev reflections (MAR) \cite{MAR}. Thus, dissipationless d.c. transport at $V=0$ and dissipative pair-assisted quasiparticle transport at $V\neq 0$ are well-separated phenomena.

Recently, set-ups involving three superconductors coupled by a common link have been proposed theoretically \cite{CUEVAS-POTHIER,HOUZET-DUHOT,HOUZET-SAMUELSSON,VINOKUR,FREYN,JONCKHEERE} and recently achieved experimentally \cite{NANOSQUID,LEFLOCH-NOISE,PFEFFER}. Transport is then governed by two (instead of one) phase or voltage variables. Coherent mechanisms coupling the three superconductors altogether become possible and provide new d.c. channels which involve correlated motion of Cooper pairs. The lowest order processes correspond to quartets, e.g, two pairs exiting simultaneously from a reservoir and crossing the bijunction, one towards each of the other reservoirs \cite{FREYN,JONCKHEERE}. This process is deeply related to so-called Cooper pair splitting that occurs in an hybrid $NSN$ bijunction. Cooper pairs are split from $S$ by so-called Crossed Andreev Reflection, and two energy- and spin- entangled electrons are sent in the two separated normal metal reservoirs \cite{BYERS,TORRES,DEUTSCHER,FALCI,RECHER,LESOVIK,MELIN,BECKMANN,RUSSO,CHANDRASEKHAR,TAKIS,BASEL,HEIBLUM}. When those reservoirs are instead superconducting, such a process should be duplicated in order to produce two quasiparticles in each of the side superconductors, that recombine into Cooper pairs \cite{FREYN,JONCKHEERE,PFEFFER}. The basic process thus involves {\it four} fermions, and it originates from splitting {\it two} Cooper pairs from a given superconductor. It involves four (instead of two) Andreev reflections within the weak link $N$, and it exchanges opposite-spin electrons of the two split Cooper pairs. 

\begin{figure}[h]
\begin{center}
\includegraphics[width=9pc]{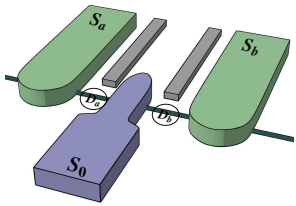}\hspace{6pc}%
\begin{minipage}[b]{14pc}\caption{\label{Bijunction}\small{A Josephson bijunction. Superconductors $S_{a,b}$
 are biased at voltages $V_{a,b}$, $S_0$ is grounded. The distance between
  the two quantum dots is comparable to the
  coherence length. Gates (in grey online) are figured.}}
\end{minipage}
\end{center}
\end{figure}

Here we review the present theory of quartet and multipair coherent transport, and summarize a recent experiment where transport signatures of quartets have been obtained in a triterminal $Al-Cu$ bijunction\cite{PFEFFER}.  Section 2 recalls the adiabatic argument for multipair transport. Section 3 presents theoretical results for a double dot bijunction. Section 4 summarizes the recent experiment and the quartet interpretation. 

\section{The adiabatic limit of multipair transport}
Let us consider three superconductors $S_{0,a,b}$ with voltages $V_0=0, V_a$ and $V_b$, connected by a short metallic link or a quantum (double) dot (Figure \ref{Bijunction}). Defining the phases $\varphi_0=0,\varphi_a$, $\varphi_b$, the total free energy at equilibrium ($V_{a,b}=0$) can be written as a doubly periodic function $F(\varphi_a,\varphi_b)$. Fourier transforming and expressing the current in one lead (say, $S_a$), one obtains:
\begin{equation}
\label{CurrentFourier}
I_a\,=\,\frac{2e}{\hbar}\frac{\partial F}{\partial \varphi_a}\,=\,\sum_{n,m}\,I_a(n,m)\,\sin(n\varphi_a+m\varphi_b)
\end{equation}
 At nonzero voltages, one writes:
\begin{equation}
\label{Adiabatic}
\varphi_a\,=\,\varphi_{0a}+\frac{2e}{\hbar}V_a\,t\,,\;\;\;\varphi_b\,=\,\varphi_{0b}+\frac{2e}{\hbar}V_b\,t
\end{equation}
where the phases at the origin of time should not be omitted. Within the adiabatic approximation, one plugs the time-dependent phases into the $I_a(\varphi_a,\varphi_b)$ relation \cite{TINKHAM}. Thus, if $V_{a,b}$ satisfy $nV_a+mV_b=0$, only the harmonics $\sin p(n\varphi_a+m\varphi_b)$ of $I_a$ have a nonzero static value, and a d.c. current flows in $S_a$, as well as in $S_0$ and $S_b$. This current involves $n$ ($m$) pairs crossing altogether to $S_a$($S_b$), and $n+m$ pairs coming from $S_0$. The simplest of these multipair processes involves quartets, e.g. $n=m=1$. It is important to notice that the lowest-order Josephson processes taking one single pair from one to the other superconductor average to zero. 

From Equations (\ref{CurrentFourier},\ref{Adiabatic}), one finds that a d.c. multipair current in $S_a$ appears:
\begin{equation}
\label{Multipair}
I_a^{d.c.}(\varphi_{0a},\varphi_{0b})\,=\,\,\sum_{p}\,I_{a,max}(p)\,\sin p(n\varphi_{0a}+m\varphi_{0b})
\end{equation}
This shows the relevance of the phase combination $(n\varphi_{0a}+m\varphi_{0b})$ which controls coherent d.c. transport by multipair processes. This phase can be emulated either by imposing suitable fluxes in a circuit containing the bijunction, or by sending a d.c. current in the bijunction. The first method is demanding as it implies the control of two phase variables. The second one is somewhat similar to driving a current into a Josephson junction and exploring its zero-voltage branch. An equivalent in the bijunction of this zero-voltage branch is a  $nV_a+mV_b=0$ line in the $(V_a,V_b)$ plane. 

The above simple calculation is valid only for tunneling barriers and voltages well below the gap. Yet, in this case, the quartet process which is a second order process might be weak. On the contrary, if the bijunction is metallic and has a high transparency, quartet transport can be strong and even resonant. In this case, a full nonequilibrium calculation should be performed. In what follows, we explore two situations, first, a bijunction made of a double quantum dot, achievable with a carbon nanotoube or a nanowire, and second, with disordered long metallic junctions.

\section{Quartet and multipair transport in a double quantum dot bijunction} 
Recent progress on Cooper pair splitting has arisen with the fabrication of hybrid $N_aD_aS_0D_bN_b$ bijunctions where the dots $D$ are created within a single wall carbon nanotube \cite{TAKIS} or a semiconducting nanowire \cite{BASEL,HEIBLUM}. Evidence for Cooper pair splitting comes out from measurement of the nonlocal conductance. Consider now a bijunction $S_aD_aS_0D_bS_b$ with a similar geometry, but with all leads being superconducting. For sake of simplicity, one considers one single electronic level $\varepsilon_{a,b}$ in each dot. The coupling between the dots and the superconductors is supposed to be symmetric and described by the level broadening $\Gamma=\pi \nu (0) t^2$ where $\nu$ is the normal metal density of states in the superconductors and $t$ is a dot-superconductor hopping parameter. The adiabatic argument of Section 2 suggests that the multipair current depends on the phase combination $n\varphi_{0a}+m\varphi_{0b}$. 
The calculation of the current-phase characteristics is performed using nonequilibrium Keldysh Green's functions \cite{JONCKHEERE}. This yields the solution as an effective partition function for the dot ``dressed'' by  tunneling into the superconductors. This approach is nonperturbative and provides exact results. Details about the formalism can be found in Ref. \cite{CHEVALLIER}. The parameters of the calculation (in units of the gap) are : the dot energies $\varepsilon_{a,b}$, the coupling $\Gamma$, and the voltage $V$. Notice that the calculation gives the full current, which for a given commensurate combination of voltages comprises not only the coherent d.c. multipair current, but also MAR quasiparticle currents.

\begin{figure}[h]
\begin{center}
\includegraphics[width=20pc]{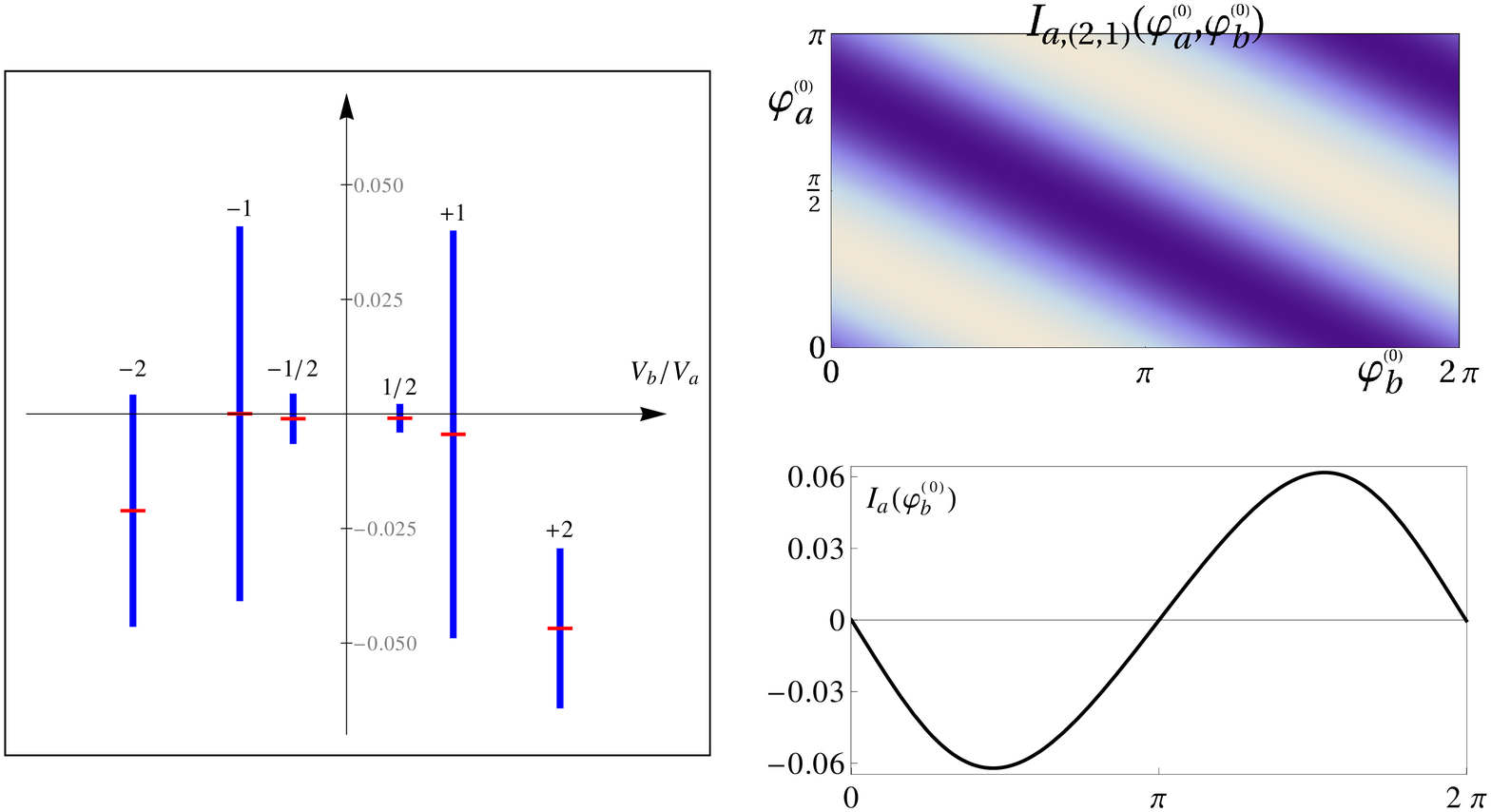}\hspace{2pc}%
\includegraphics[width=10pc]{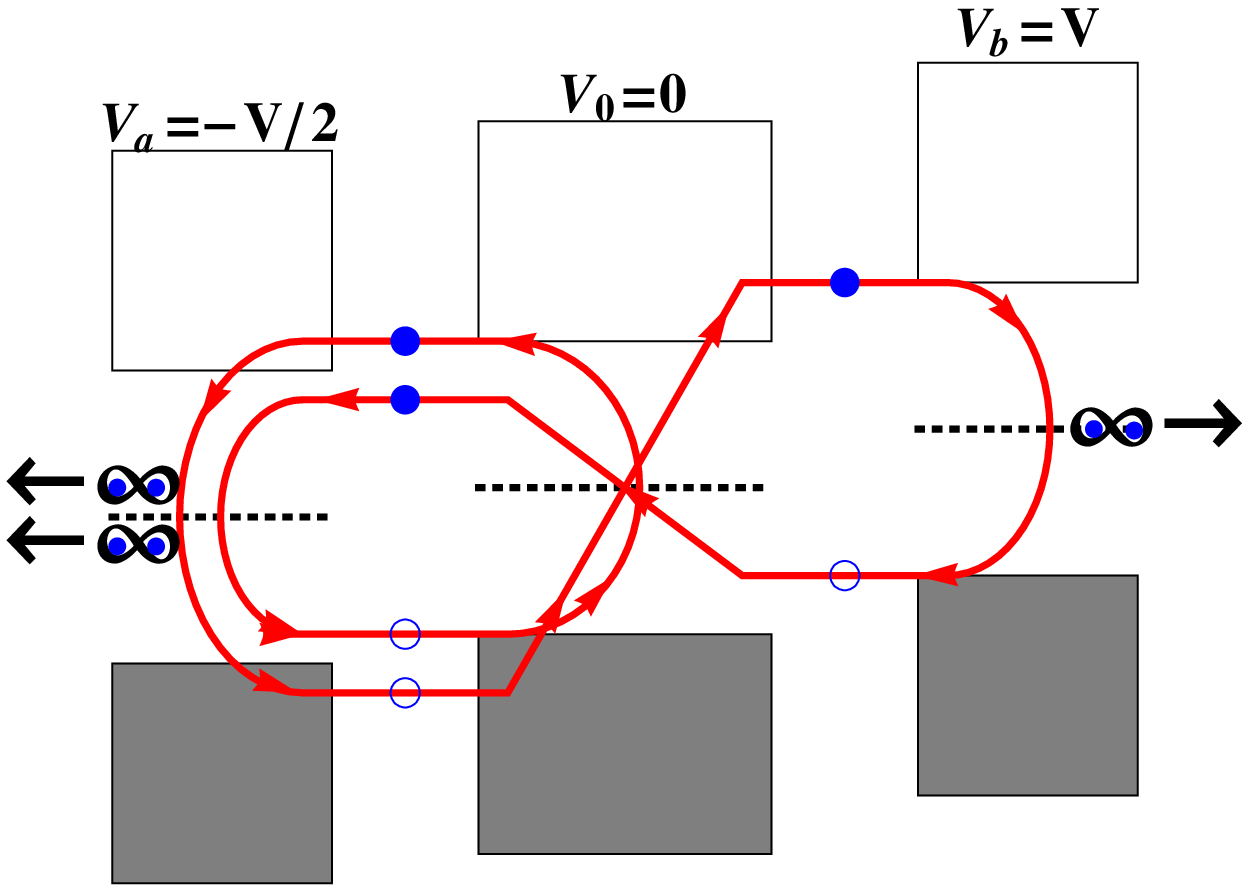}
\caption{\label{fig:metallic}(color online)\small{``Broad'' dots regime:  $|\varepsilon_{a,b}|=6\Delta$, $\Gamma_{a,b}=4\Delta$. 
    Left:  phase-dispersion (in blue online) of the d.c. current for the main resonances (with $|n|+|m| \leq 3$), 
    centered around the values of the phase-independent  quasiparticle current (small horizontal bars).
   Horizontal axis is $V_a/V_b$, with $V_b/\Delta=0.3$.
   Upper middle: Current $\langle I_a \rangle$, for the resonance $2 V_a + V_b =0$, as a function of the 
   phases $\varphi_a^{(0)}$ and $\varphi_b^{(0)}$, showing the dependence in $2 \varphi_{a}^{(0)} + \varphi_{b}^{(0)}$.
   Lower middle: Current-phase relation $I_a(\varphi_b^{(0)})$ at $\varphi_a^{(0)}=0$ for the quartet resonance $V_a+V_b=0$,
   which shows the $\pi$-phase behavior. Right: Energy diagram for the
  $S_aD_aS_0D_bS_b$ bijunction, with a higher order diagram associated with a
  ``sextet'' current with 3 pairs emitted from $S_0$, 2 into $S_a$ and 1 into $S_b$, with $V_a=-V_b/2$.}}
   \end{center}
\end{figure}

\subsection{Open dot regime}
One first considers a regime in which each dot mimics a metallic junction, achieved by placing energy levels out of resonance $|\varepsilon_{a,b}| > \Delta$, with large couplings $\Gamma >\Delta$. The results for the largest resonances are represented on Figure \ref{fig:metallic} and compared to the phase-independent part of the quasiparticle current. One sees that the multipair features appear as strong resonances. The quartet dispersion has a negative slope at the origin and is thus a $\pi$-junction in terms of the ``quartet phase'' $\varphi_a+\varphi_b$. This is due to the exchange process occurring as two split Cooper pairs are emitted by $S_0$ and recombine into one pair in $S_a$ and one in $S_b$. 

\subsection{Closed and nonresonant dot regime}
\begin{figure}[h]
\begin{center}
\begin{minipage}{14pc}
\includegraphics[width=1.1\textwidth]{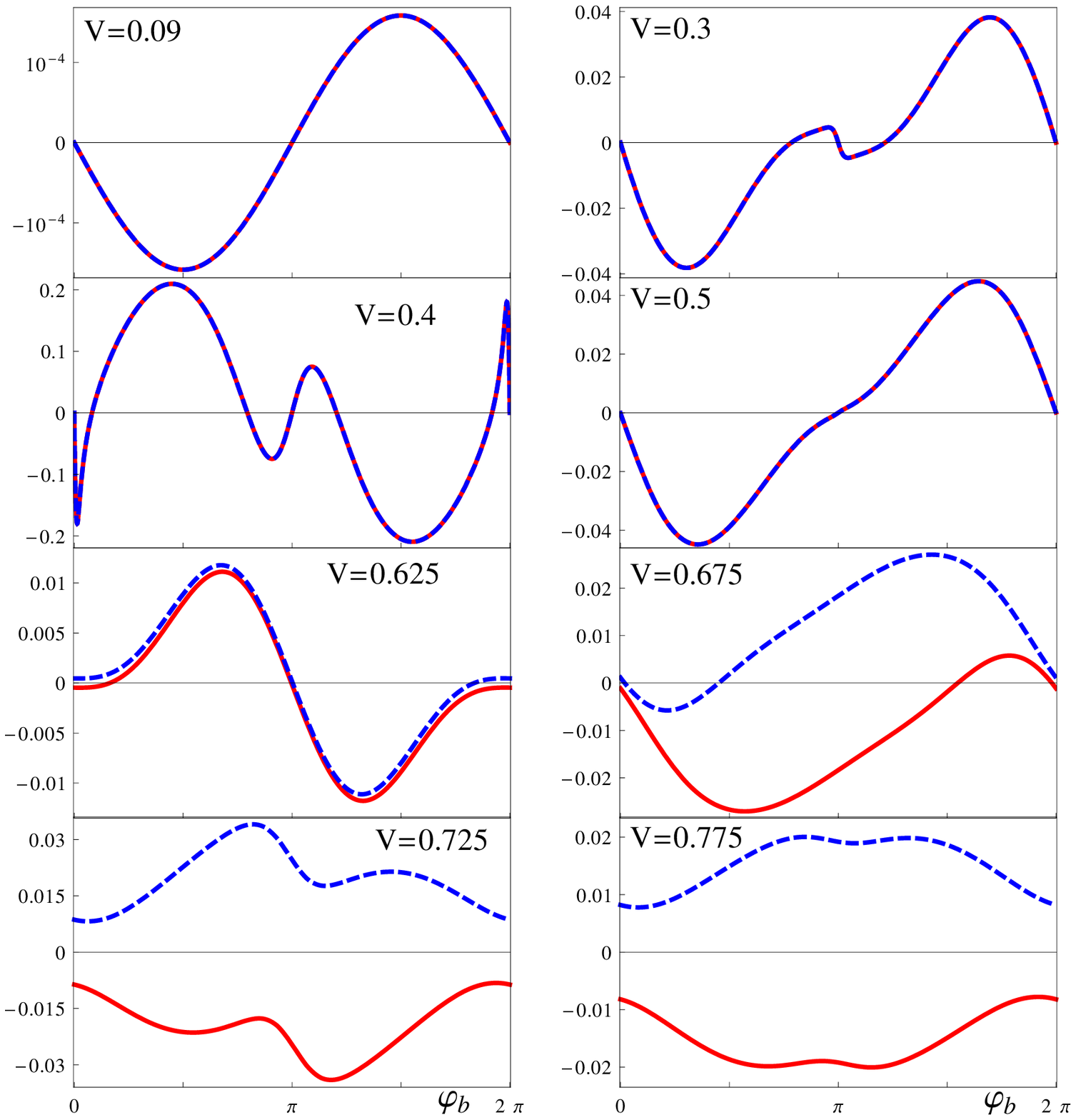}
\caption{\small{(color online) Current-phase relations $I_a(\varphi_b^{(0)})$ (red, full curve)
 and $I_b(\varphi_b^{(0)})$ (blue, dashed curve) in the quartet configuration 
$V_a = - V_b$, for the resonant dots regime: $\varepsilon_a=-\varepsilon_b=0.4 \Delta$, $\Gamma=0.1 \Delta$, $V=0.09-0.775 \Delta$.
 Note the different $y$ scales in the different panels.
\label{fig:criticalcurrent}}}
\end{minipage}\hspace{6pc}%
\begin{minipage}{14pc}
\includegraphics[width=4.cm]{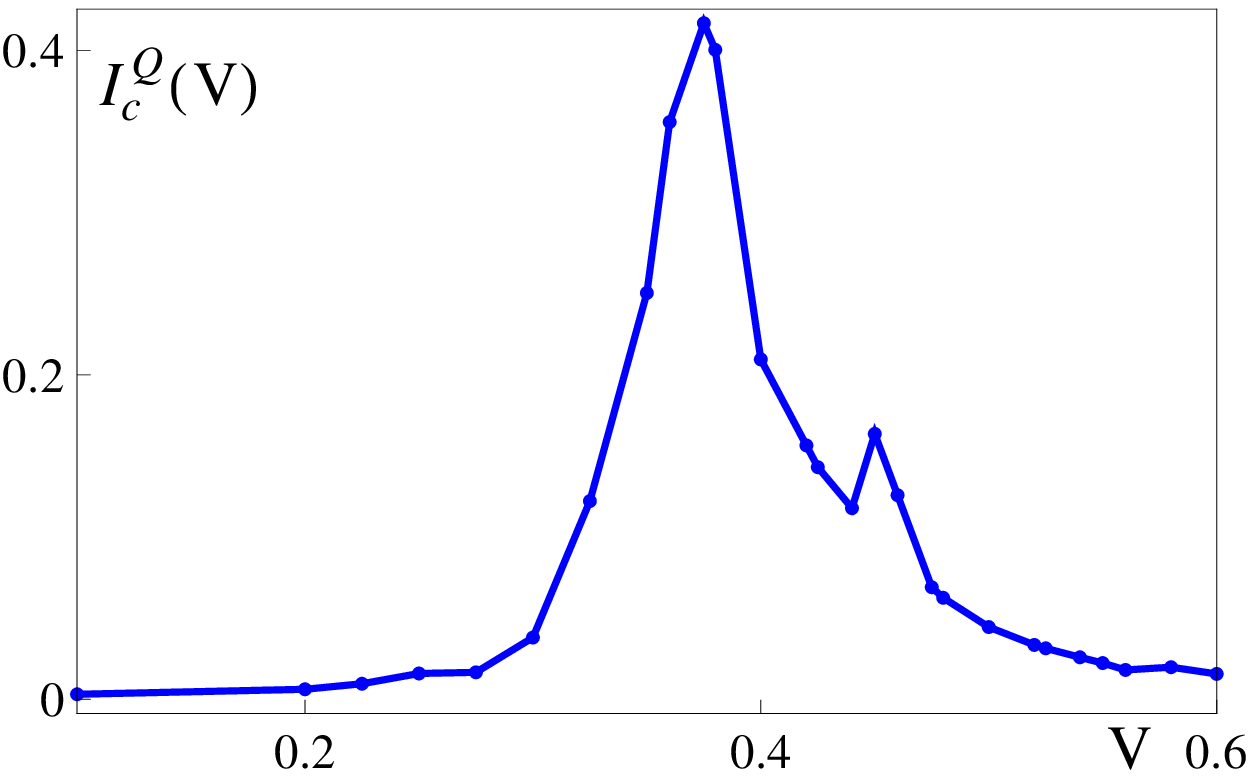}
\hspace{.7cm}
\includegraphics[width=0.7\textwidth]{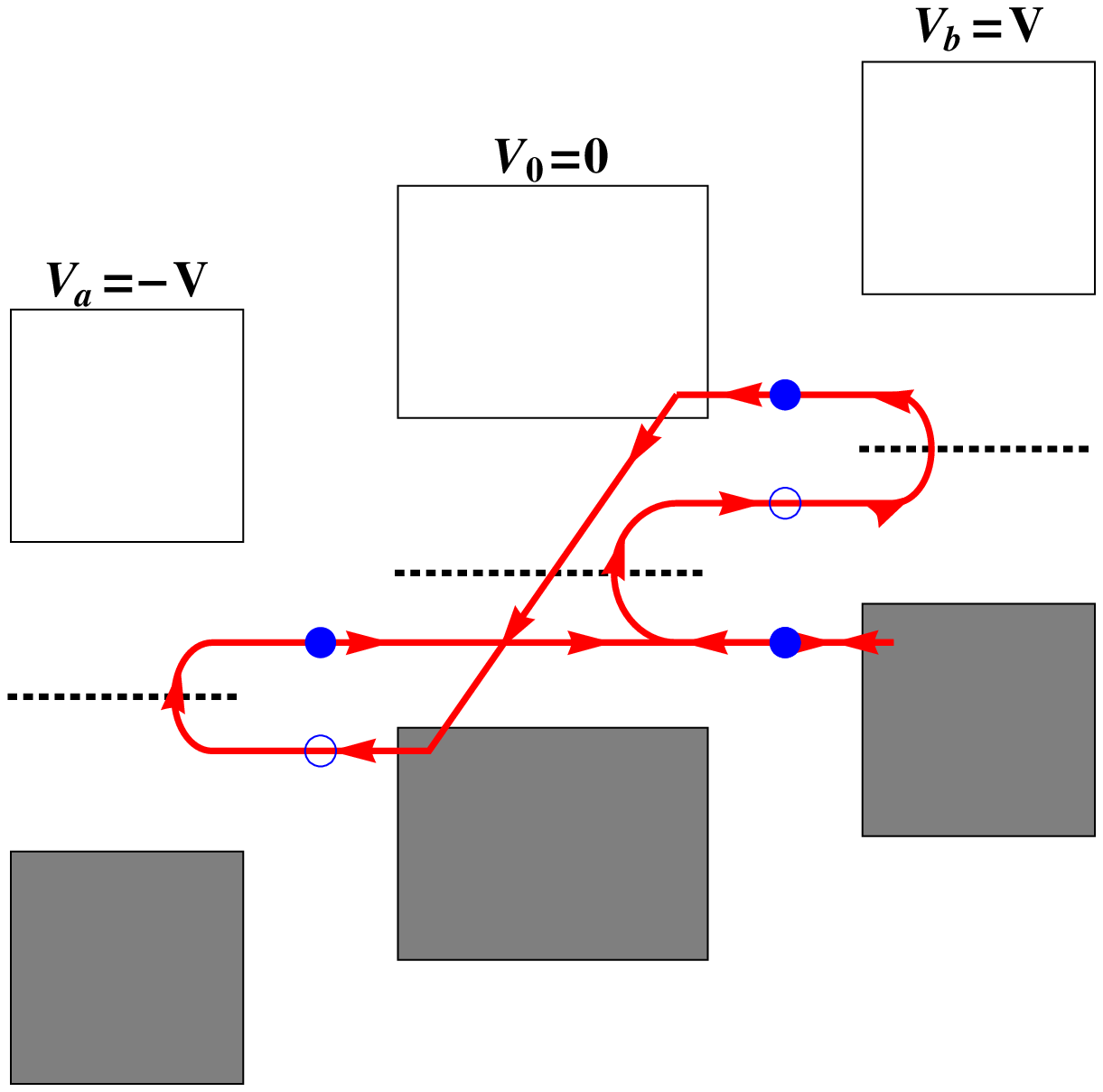}
\caption{\small{(color online)
	Upper panel: critical current $I_c^{Q}$ a as function of the voltage in the quartet configuration, for resonant dots
	 (same parameters as in Figure \ref{fig:criticalcurrent}. 
	 Lower panel: Lowest order diagram contributing to the phase-dependent MAR current.}}
\label{fig:ProcMixtes}
\end{minipage} 
\end{center}
\end{figure}

Consider now small coupling and nonresonant dots, e.g. $\Gamma < |\varepsilon_{a,b}| < \Delta$. Figure \ref{fig:criticalcurrent} shows the current-phase profiles for various values of $V$, increasing from $0.09\Delta$ to $0.8 \Delta$. The maximum current increases by several orders of magnitude and is maximum when $V \sim  |\varepsilon_{a,b}|$. Moreover, above the maximum, currents $I_a$ and $I_b$ start to depart from each other, signalling the onset of a strong MAR quasiparticle current. Remarkably, this MAR current is phase-dependent, a specific feature of the resonant quartet behaviour, which can be ascribed to interferences between various amplitudes within the bijunction (Figure \ref{fig:ProcMixtes}). 

This calculation validates the existence of quartet and higher-order resonances featuring a d.c. current of correlated Cooper pairs. These currents are functions of a well-defined combination of the phases, that becomes a relevant and observable variable. Phase coherence is thus established between the three superconductors, despite the presence of  voltage biases. One should underline that instead of Andreev bound states forming in a junction at equilibrium, here the d.c. Josephson-like multipair resonances are due to Andreev {\it resonances}, and they coexist with dissipative MAR channels. This coexistence is a novel feature in superconductivity. As a corollary, the maximum "critical" quartet current depends on the applied voltage, in a dramatic way in the example shown above (Section 3.2).

\section{Quartet anomalies in metallic bijunctions}
\begin{figure}[h]
\begin{center}
\includegraphics[width=16pc]{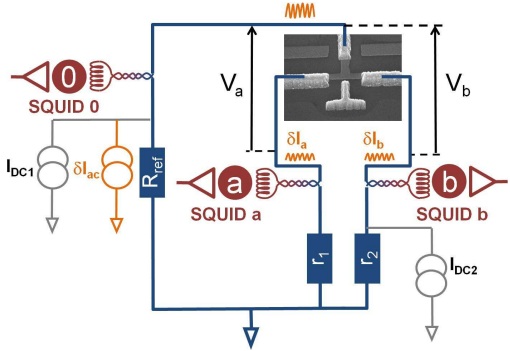}\hspace{4pc}%
\begin{minipage}[b]{14pc}\caption{\label{Measurement}\small{Experimental set-up \cite{COUPIAC}. The three macroscopic resistors have low resistance values ($\simeq 0.1$ $\Omega$) allowing voltage-biasing of the samples. The SEM image shows a bijunction sample with a T-shape geometry.}}
\end{minipage}
\end{center}
\end{figure}

Copper and Aluminum bijunctions have been fabricated \cite{PFEFFER} by shadow mask evaporation technique \cite{DUBOS}, leading to highly transparent and uniform SN interfaces. They form three-terminal junctions with a T-shape normal conductor connecting three superconducting electrodes $S_0, S_a$ and $S_b$ (Figure \ref{Measurement}). The width and length of the normal metal are about $0.6 \,\mu m$ and $1 \,\mu m $. Using a diffusion constant for copper $D = 100 \, cm^2/s$, one gets a Thouless energy $E_{Th} = \hbar D/L^2 = 7 \, \mu eV$. The superconducting Al energy gap is $\Delta = 170 \,\mu eV$. These Josephson junctions are thus in the long junction regime defined as the Thouless energy $E_{Th}$ being much smaller than the gap $\Delta$, or equivalently a junction length $L$ larger than the superconductor coherence length $\xi_s=\sqrt{\hbar D/\Delta}$. The diffusion time is $\tau_D=L^2/D \simeq 0.1 ns$ is much smaller than the inelastic time $\tau_{in}\simeq 1 \,ns$ at $100 \, mK$.

\begin{figure}[h]
\begin{center}
\begin{minipage}{14pc}
\includegraphics[width=1.1\textwidth]{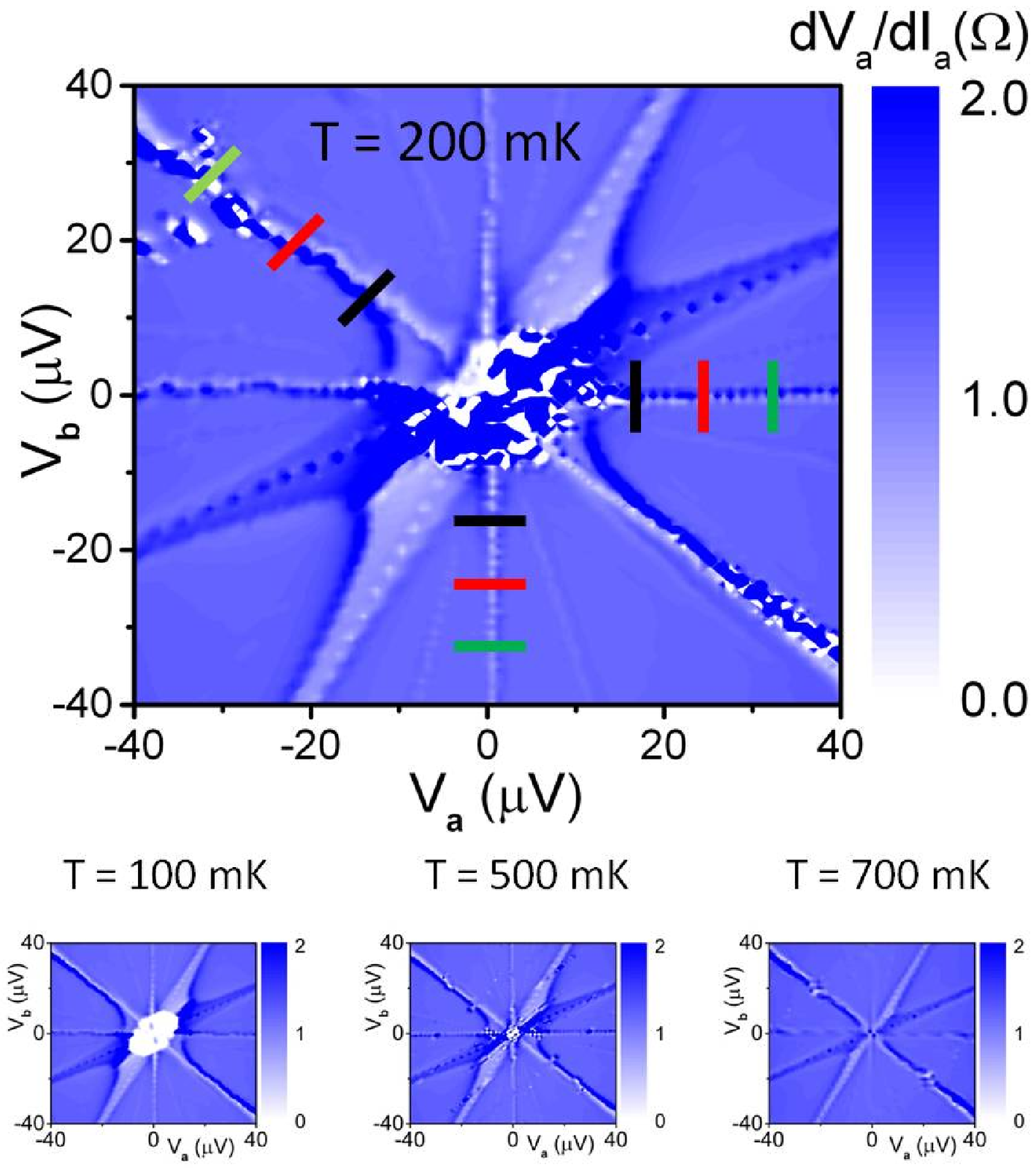}
\caption{\small{Differential resistance $R_{diff,a}$ of a T-shape junction in the ($V_a,V_b$) plane for various temperatures.}}
\label{Tri}
\end{minipage}\hspace{2pc}%
\begin{minipage}{14pc}
\includegraphics[width=14pc]{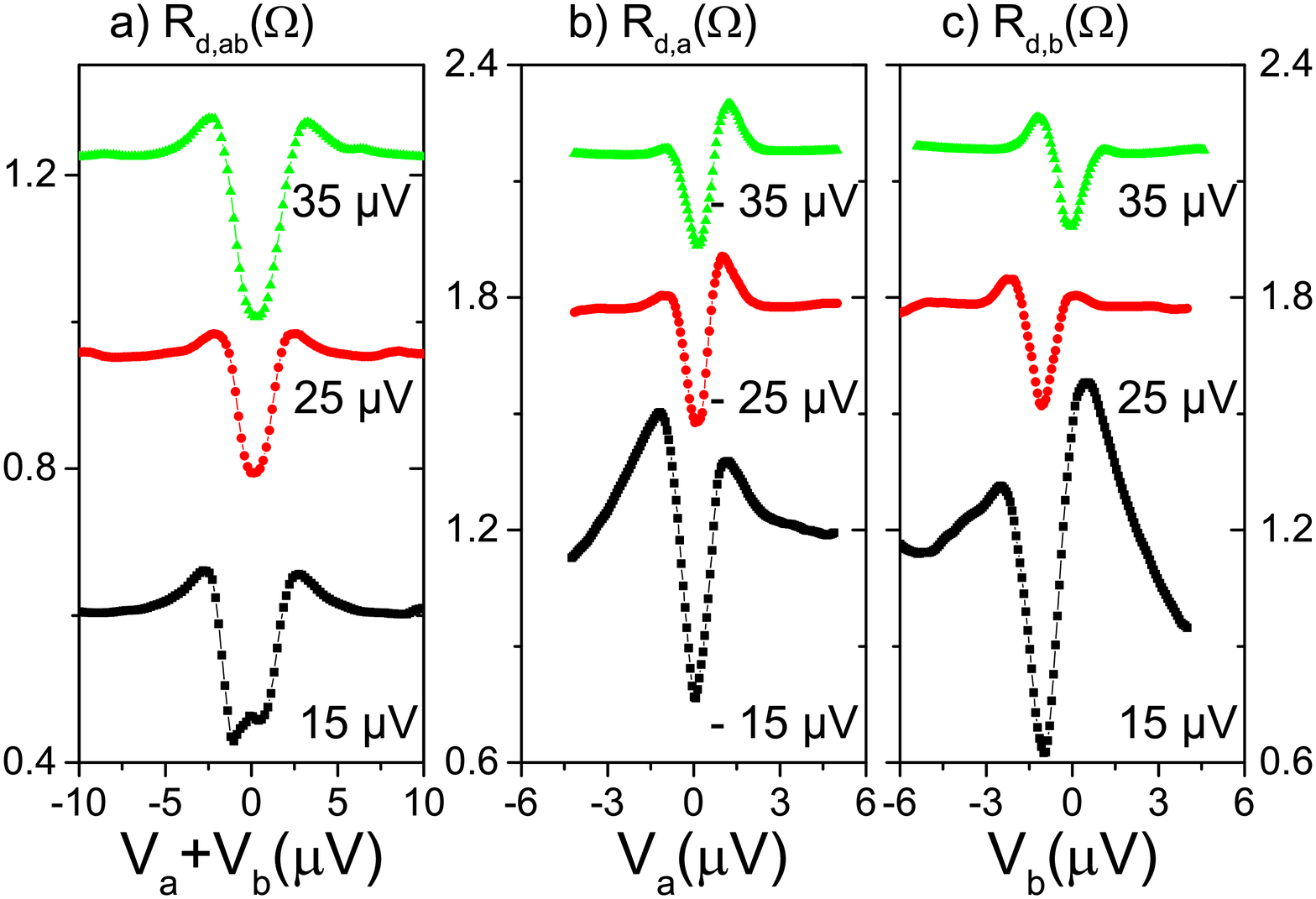}
\caption{\small{Line-traces at various values of the applied voltage of a) the differential resistance of the full sample as a function of the voltage $V_a+V_b$, for several values of  $(V_a-V_b)/2$, b) c) the differential resistance of branch $a$ vs $V_a$ ($V_b$) for various values of $V_b$ ($V_a$).}}
\label{Traces}
\end{minipage}
\end{center}
\end{figure}

Three-terminal differential resistances were measured using a specific experimental set-up including 3 SQUIDs as current amplifiers (Figure \ref{Measurement}) \cite{COUPIAC}. The differential resistance $\frac{\partial V_a}{\partial I_a}$ is plotted as a 2D map in the $(V_a,V_b)$ plane (Figure \ref{Tri}). First, it displays anomalies along the axis, that correspond to the "direct" Josephson effect  between $S_0$ and $S_a$ or $S_b$ (the one between $S_a$ and $S_b$ is not visible due to the location of the ac lock-in).  
Second, clear features appear at voltage values such as $V_i+V_j-2V_k=0$, where $(i,j,k)=(a,b,0), (b,0,a), (0,a,b)$, corresponding to quartets emitted respectively by $S_0, S_a, S_b$. The transversal $I(V)$ traces shown in Figure \ref{Traces} display comparable shape and amplitude with respect to the Josephson features obtained at $V_a=0$ or $V_b=0$. 

\begin{figure}[h]
\begin{center}
\includegraphics[width=16pc]{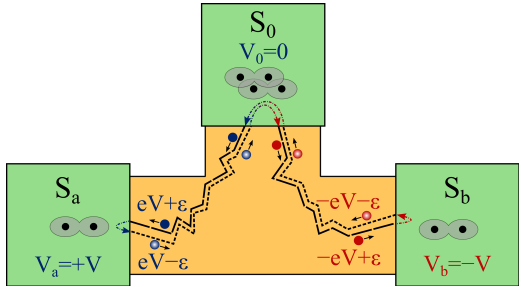}\hspace{2pc}%
\begin{minipage}[b]{14pc}\caption{\label{Quartet}\small{Schematic diagram for $Q_0$ quartet production, from $S_0$ to $S_a$ and $S_b$. Two Cooper pairs are split simultaneously at $S_0$ with one electron of each pair diffusing towards $S_a$ and $S_b$ where, under the appropriate energy condition ($V_a=-V_b$), they recombine to create two separated Cooper pairs.}}
\end{minipage}
\end{center}
\end{figure}

An explanation in terms of the synchronization of ac Josephson oscillations in junctions $S_0S_a$ and $S_0S_b$ \cite{NERENBERG,ARGAMAN} is unlikely, since in the experimental regime  where $eV$ is as large as $8 E_{Th} $, ac Josephson oscillations should be strongly decreased. In addition, the amplitude of the anomalies is nearly independent of the voltage value, pointing towards a fully coherent phenomenon. Figure \ref{Quartet} indeed shows why the quartet mechanism is immune against the decoherence brought by such voltage in a single Josephson junction. Two Andreev reflections in $S_0$ produce two splitted Cooper pairs, materializing in electrons and holes at energies $eV\pm \varepsilon$ in the $a-$ branch, and  at energies $-eV\pm \varepsilon$ in the $b-$ branch. Provided $\varepsilon < E_{Th}$, such energies allow recombination by Andreev reflection at $S_a$ and $S_b$, without any loss of coherence introduced by the voltage. This coherent process has the structure of a MAR process, but in the present case it is energy conserving and it provides a Josephson-like quartet channel.

A perturbative non-equilibrium Green's function \cite{PFEFFER} calculations gives the following estimate for the maximum quartet current:

\begin{equation}
\label{eq:IQ}
eI_Q \sim -G_{CAR} E_{Th} \sin(\varphi_{a0}+\varphi_{b0})
,
\end{equation}
where the characteristic phase dependence of the quartet mode stems from the four involved Andreev reflections, one at $S_a$, one at $S_b$ and two at $S_0$. The conductance $G_{CAR}$ is the Crossed Andreev conductance of a $N_aNS_0NN_b$ structure in which the electrodes $S_a$ and $S_b$ are in the normal state and at voltages $\pm V$. By ``CAR conductance'' we
mean the conductance formally associated to a process in which an electron from $N_a$, at energy $eV$ is transmitted as a hole in $N_b$ at energy $-eV$, and a Cooper pair is therefore split from $S_0$ \cite{MELIN}. This formula is similar to the one for a SNS junction where the {\it pair} critical current $eI_c \sim G_{N} E_{Th}$ is related to the single particle normal conductance and the Thouless energy sets the coherence of Andreev reflection. 

The CAR conductance can be evaluated as $G_{CAR} \sim \frac{G_{Na}G_{Nb}}{G_{0}(\xi_s)}$, 
where $G_{Na,b}$ is the conductance within each normal branch of the bijunction, and $G_0(\xi_s)=(2e^2/h) {\cal N} (l_e/\xi)$ is the normal-state conductance of a region of size $\xi$ of the superconductor $S_0$ (${\cal N}$ is the channel number).
This shows that the ratio between the quartet maximum current at a bias $V$ and the single junction critical current at zero bias is $I_{Qmax}(V)/I_c(0) \sim G_{CAR}/G_N \sim G_N/G_0(\xi_s)$, which is not necessarily small. Based on measured sample parameters, we estimate this ratio to $\sim 0.1-0.5$, in fair agreement with the experiment. Notice that if $eV \ll \Delta$, then $G_{CAR}$ as well as $I_{Q,max}$ do not decrease with $V$, in agreement with the present experiment.

\section{Conclusion}
The theory and the experiment reported here open the way to a new kind of Josephson devices. Phase coherence is possible between three superconductors biased at suitable voltages, and open new channels for transport, made of correlated Cooper pairs. Such phase coherence remains to be demonstrated by interference experiments. Beyond the understanding and probing of the microscopic properties of such bijunctions (made of quantum dots or metals), future prospects include the exploration of new entanglement properties, and the production of phase-correlated microwave fields. 

\subsection*{Acknowledgments}
This work has been partially funded by ANR-NanoQuartet (ANR12BS1000701). We acknowledge the Nanoscience Foundation for the PhD grant of A. H. Pfeffer and the NanoFab facility at Institut N\'eel-CNRS for sample fabrication.

\end{document}